\documentclass{ws-procs975x65}
\def\beq{\begin{equation}}
\def\eeq{\end{equation}}

\newcommand{\be}{\begin{equation}}
\newcommand{\ee}{\end{equation}}
\newcommand{\ba}{\begin{eqnarray}}
\newcommand{\ea}{\end{eqnarray}}

\newcommand{\lp}{\left(}
\newcommand{\rp}{\right)}
\newcommand{\lb}{\left[}
\newcommand{\rb}{\right]}
\begin{document}

\title{Novel couplings between nonmetricity and matter}

\author{Tiberiu Harko}
\address{Department of Physics, Babes-Bolyai University,\\ Kogalniceanu Street,
Cluj-Napoca 400084, Romania, and\\
Department of Mathematics, University College London,\\Gower Street, London
WC1E 6BT, United Kingdom\\
E-mail: t.harko@ucl.ac.uk}

\author{Tomi S. Koivisto}
\address{Nordita, KTH Royal Institute of Technology and Stockholm University,\\ Roslagstullsbacken 23, 10691 Stockholm, Sweden\\
E-mail: timoko@kth.se}

\author{Gonzalo J. Olmo}
\address{Departamento de F\'{i}sica Te\'{o}rica and IFIC, Centro Mixto Universidad de Valencia - CSIC.
Universidad de Valencia, Burjassot-46100, Valencia, Spain\\
E-mail: gonzalo.olmo@uv.es}

\author{Francisco S.N. Lobo$^*$ and Diego Rubiera-Garcia$^{**}$}
\address{Instituto de Astrof\'{\i}sica e Ci\^{e}ncias do Espa\c{c}o, \\Faculdade de
Ci\^encias da Universidade de Lisboa, \\Edif\'{\i}cio C8, Campo Grande,
P-1749-016 Lisbon, Portugal\\
$^*$E-mail: fslobo@fc.ul.pt\\
$^{**}$E-mail: drgarcia@fc.ul.pt}

%%%%%%%%%%%%%%%%%%%%%%%%%%%%%%%%%%%%%%%%%%%%%%%%%%%%%%%%%%%%%%%%%%%%%%%%%%%%%%%%%%%%

\begin{abstract}
We present a novel theory of gravity, namely, an extension of symmetric teleparallel gravity. This is done by introducing a new class of theories where the nonmetricity $Q$ is coupled nonminimally to the matter Lagrangian. This nonminimal coupling entails the nonconservation of the energy-momentum tensor, and consequently the appearance of an extra force. We also present several cosmological applications.
\end{abstract}


\bodymatter

%%%%%%%%%%%%%%%%% now a standard article style for the most part

\section{Introduction}

The fact that general relativity (GR) is facing so many challenges, namely, (i) the difficulty in explaining particular observations, (ii) the incompatibility with other well established theories, (iii) and the lack of uniqueness, may be indicative of a need for new gravitational physics. Thus, a promising approach is to assume that at large scales GR breaks down, and a more general action describes the gravitational field\cite{modgrav,modgrav2,modgrav3,Capozziello:2011et,Lobo:2008sg}. The physical motivations for these modifications of gravity include the following: (i) the possibility of a more realistic representation of the gravitational fields near curvature singularities; (ii) and to create some first order approximation for the quantum theory of gravitational fields.

An interesting model that is related to this work is the nonminimal curvature-matter coupling \cite{Goenner,Koivisto:2005yk,Bertolami:2007gv,BookHarkoLobo}, in which the Lagrangian density is of the form $L \sim f_1(R)+\left[1+\lambda f_2(R)\right]{L}_{m}$,
%\begin{equation}
%S=\int \left\{\frac{1}{2}f_1(R)+\left[1+\lambda f_2(R)\right]{
%L}_{m}\right\} \sqrt{-g}\;d^{4}x~,
%\label{action2}
%\end{equation}
where $f_i(R)$ (with $i=1,2$) are arbitrary functions of the Ricci scalar $R$ and ${L}_{m}$ is the matter Lagrangian density. Applications have been explored such as in an effective dark energy or dark matter \cite{BookHarkoLobo,Allemandi:2005qs,Nojiri:2004bi,Harko:2014gwa}. An interesting feature of these theories is the non-conservation of the energy-momentum tensor, so that the coupling between the matter and the higher derivative curvature terms describes an exchange of energy and momentum between both.

Indeed, it is possible to tackle gravitation through several approaches, namely, with the metric formalism\cite{modgrav,modgrav2,modgrav3}, which consists on varying the action with respect to the metric and setting the Levi-Civita connection, or through the metric-affine formalism \cite{Olmo:2011uz}, where the metric and the affine connection are regarded as independent variables, or even through a hybrid approach\cite{Harko:2011nh,Capozziello:2015lza,BookHarkoLobo}. 
Note that the metric $g_{\mu\nu}$ may be thought of as a generalization of the gravitational potential and is used to define notions such as distances, volumes and angles. On the other hand, the affine connection $\Gamma^{\mu}{}_{\alpha\beta}$ defines parallel transport and covariant derivatives.

In fact, a basic result in differential geometry is that the general affine connection may be decomposed into the following 3 independent components:
\begin{equation}
	\label{Connection decomposition}
	\Gamma^{\lambda}{}_{\mu\nu} =
	\left\lbrace {}^{\lambda}_{\phantom{\alpha}\mu\nu} \right\rbrace +
	K^{\lambda}{}_{\mu\nu}+
	L^{\lambda}{}_{\mu\nu} \,,
\end{equation}
where the first term is the Levi-Civita connection of the metric $g_{\mu\nu}$, defined as:
\begin{equation}
	\label{christoffel}
	\left\lbrace {}^{\lambda}_{\phantom{\alpha}\mu\nu} \right\rbrace \equiv \frac{1}{2} g^{\lambda \beta} \left( \partial_{\mu} g_{\beta\nu} + \partial_{\nu} g_{\beta\mu} - \partial_{\beta} g_{\mu\nu} \right) \,.
\end{equation}
The second term $K^{\lambda}_{\phantom{\alpha}\mu\nu}$ is the contortion, given by
\begin{equation}
	\label{Contortion}	
K^{\lambda}{}_{\mu\nu} \equiv \frac{1}{2} T^{\lambda}{}_{\mu\nu}+T_{(\mu}{}^{\lambda}{}_{\nu)} \,,
\end{equation}
with the torsion tensor defined as $T^{\lambda}{}_{\mu\nu}\equiv 2 \Gamma^{\lambda}{}_{[\mu\nu]}  $.
The third term is the disformation
\begin{equation}
	\label{Disformation}
	L^{\lambda}{}_{\mu\nu} \equiv \frac{1}{2} g^{\lambda \beta} \left( -Q_{\mu \beta\nu}-Q_{\nu \beta\mu}+Q_{\beta \mu \nu} \right)  \,,
\end{equation}
which is defined in terms of the nonmetricity tensor, given by $Q_{\rho \mu \nu} \equiv \nabla_{\rho} g_{\mu\nu}$.

This implies that by making assumptions on the affine connection, one is essentially specifying a metric-affine geometry \cite{Jarv:2018bgs}. For instance, the standard formulation of GR assumes a Levi-Civita connection, which implies vanishing torsion and nonmetricity, while its teleparallel equivalent (TEGR), uses the Weitzenb\"{o}ck connection, implying zero curvature and nonmetricity \cite{Maluf:2013gaa}. More recently, the symmetric teleparallel theories of gravity were analysed in \cite{BeltranJimenez:2017tkd,Conroy:2017yln,Koivisto:2018aip,BeltranJimenez:2018vdo}, which possess remarkable features.
Here, we present an extension of the symmetric teleparallel gravity, by introducing a new class of theories where a general function of the nonmetricity $Q$ is coupled nonminimally to the matter Lagrangian, in the framework of the metric-affine formalism\cite{Harko:2018gxr}. 

This work is outlined in the following manner: In Section \ref{secII}, we present the formalism of an extended symmetric teleparallel equivalent of general relativity. In Section \ref{secIII}, we consider some cosmological applications and conclude in Section \ref{secIV}.

\section{Nonmetricity-matter coupling}\label{secII}

\subsection{Action and gravitational field equations}

Consider the action defined by two functions of the nonmetricity $Q$, given by
\begin{equation}  \label{qqm}
S = \int {\mathrm{d}}^4 x \sqrt{-g}\left[\frac{1}{2} f_1(Q) + f_2(Q)L_M\right]\,,
\end{equation}
where $L_M$ is a Lagrangian function for the matter fields.

We define the nonmetricity tensor and its two traces as follows:
\begin{equation}
Q_{\alpha \mu \nu }=\nabla _{\alpha }g_{\mu \nu }\,,\qquad Q_{\alpha
}=Q_{\alpha }{}^{\mu }{}_{\mu }\,,\qquad \tilde{Q}_{\alpha }=Q^{\mu
}{}_{\alpha \mu }\,.
\end{equation}%
It is also useful to introduce the superpotential
\begin{eqnarray}
4P^{\alpha }{}_{\mu \nu } = -Q^{\alpha }{}_{\mu \nu } + 2Q_{(\mu %
\phantom{\alpha}\nu )}^{\phantom{\mu}\alpha } - Q^{\alpha }g_{\mu \nu }  
%\\&&
-\tilde{Q}^{\alpha }g_{\mu \nu }-\delta _{(\mu }^{\alpha }Q_{\nu )}\,, \label{super}
\end{eqnarray}
where one can readily check that $Q=-Q_{\alpha \mu \nu }P^{\alpha \mu \nu
}$, which is a useful relation.

%\subsection{Gravitational field equations}

For notational simplicity, we introduce the following definitions
\begin{equation}
f=f_{1}(Q)+2f_{2}(Q)L_{M}\,,\qquad F=f_{1}^{\prime }(Q)+2f_{2}^{\prime
}(Q)L_{M}\,,  \label{f_F}
\end{equation}%
and specify the following variations
\begin{eqnarray}\label{emt}
T_{\mu \nu } =-\frac{2}{\sqrt{-g}}\frac{\delta (\sqrt{-g}{L}_{M})}{\delta
g^{\mu \nu }}\,, \qquad
H_{\lambda }{}^{\mu \nu } =-\frac{1}{2}\frac{\delta (\sqrt{-g}{L}_{M})}{%
\delta \Gamma _{\phantom{\lambda}\mu \nu }^{\lambda }}\,,
\end{eqnarray}%
%
%\begin{eqnarray}
%T_{\mu \nu } &=&-\frac{2}{\sqrt{-g}}\frac{\delta (\sqrt{-g}{L}_{M})}{\delta
%g^{\mu \nu }}\,, \label{emt} \\
%H_{\lambda }{}^{\mu \nu } &=&-\frac{1}{2}\frac{\delta (\sqrt{-g}{L}_{M})}{%
%\delta \Gamma _{\phantom{\lambda}\mu \nu }^{\lambda }}\,,
%\end{eqnarray}%
as the energy-momentum tensor and the hyper-momentum tensor density,
respectively.

Varying the action (\ref{qqm}) with respect to the metric, one obtains the
gravitational field equations given by
\begin{equation}
\frac{2}{\sqrt{-g}}\nabla_\alpha\left(\sqrt{-g}FP^\alpha{}_{\mu\nu}\right)
+ \frac{1}{2}g_{\mu\nu} f_1   + F\left( P_{\mu\alpha\beta}Q_{\nu}{}^{\alpha\beta}
-2Q_{\alpha\beta\mu}P^{\alpha\beta}{}_\nu\right) = -f_2 T_{\mu\nu} \,.\label{efe}
\end{equation}
and the variation with respect to the connection, yields the following relation
\begin{equation}
\nabla_\mu\nabla_\nu \left(\sqrt{-g} F P^{\mu\nu}{}_\alpha - f_2 H_\alpha{}%
^{\mu\nu}\right) = 0\,.  \label{cfe}
\end{equation}

\subsection{Divergence of the energy-momentum tensor}

The divergence of the energy-momentum tensor is given by:
\begin{eqnarray}
\mathcal{D}_\mu T^\mu{}_\nu +   \frac{2}{\sqrt{-g}}\nabla_\alpha \nabla_\beta H_\nu{}^{\alpha\beta} 
	&=& -\frac{2}{\sqrt{-g}f_2}\lb\lp\nabla_\alpha\nabla_\beta f_2\rp H_\nu{}^{\alpha\beta} + 2f_{2,(\alpha}\nabla_{\beta)}H_\nu{}^{\alpha\beta}\rb
	\nonumber \\
& &- \lp T^\mu{}_\nu - \delta^\mu_\nu L_M \rp \nabla_\mu \log{ f_2}\,, \label{divergence}
\end{eqnarray}
where $\mathcal{D}_\alpha$ denoted the metric covariant derivative with respect to the symbols (\ref{christoffel}). This shows that due to the coupling between the nonmetricity $Q$ and the matter fields, the matter energy-momentum tensor is no longer conserved. The first term of the right-hand-side line is due to the nonminimal coupling of the hypermomentum, and the second term is related to the nonminimal coupling of the energy-momentum tensor.

%\subsubsection{The energy and momentum balance equations}

Consider a perfect fluid, given by the energy-energy tensor, $T_{\mu \nu}=\left(\rho+p\right)u_{\mu}u_{\nu}+pg_{\mu \nu}$, so that the energy and momentum balance equations gives
\be\label{37}
\dot{\rho}+3\mathcal{H}\left(\rho+p\right)=\mathcal{S},
\ee
where $\dot\; =u_{\mu}\mathcal{D}^{\mu}$, and $\mathcal{H}=(1/3)\mathcal{D}^{\mu}u_{\mu}$. For simplicity, one can decompose the energy source term as
$\mathcal{S}=\mathcal{S}_\mathcal{T} + \mathcal{S}_\mathcal{H}$,
%\begin{equation}
%\mathcal{S}=\mathcal{S}_\mathcal{T} + \mathcal{S}_\mathcal{H}\,,
%\end{equation}
where $\mathcal{S}_\mathcal{T}$ is defined by
\be \label{s_t}
\mathcal{S}_\mathcal{T} = \lp \rho + L_M \rp \frac{\dot{f_2}}{f_2}\,,  \nonumber 
\ee
which vanishes when we adopt the Lagrangian $L_M=-\rho$, and the hypersource is given as
\begin{eqnarray}
\mathcal{S}_\mathcal{H}  =   -\frac{2}{\sqrt{-g}}u^\nu\Big[\nabla_\alpha \nabla_\beta H_\nu{}^{\alpha\beta} %\nonumber \\
 +  \frac{1}{f_2}\lp\nabla_\alpha\nabla_\beta f_2\rp H_\nu{}^{\alpha\beta} + \frac{1}{f_2}f_{2,(\alpha}\nabla_{\beta)}H_\nu{}^{\alpha\beta}\Big]\,. \nonumber 
\end{eqnarray}

%\subsubsection{Presence of an extra-force}

Note that the non-conservation of the energy-momentum tensor implies non-geodesic motion, where an extra-force $\mathcal{F}^{\lambda}$ arises due to the $Q$-matter coupling
\be\label{39}
\frac{d^2x^{\lambda}}{ds^2}+\left\lbrace {}^{\lambda}_{\phantom{\alpha}\mu\nu} \right\rbrace  u^{\mu}u^{\nu}=\mathcal{F}^{\lambda}.
\ee

The extra-force can be decomposed as
\be
\mathcal{F}^{\lambda} = -\frac{h^{\alpha \lambda}\nabla_\alpha p}{\rho+p} + \mathcal{F}^{\lambda}_{\mathcal{T}} + \mathcal{F}^{\lambda}_{\mathcal{H}}\,,
\ee
where the first term on the right-hand-side is the usual general relativistic contribution of the pressure gradient.
The extra force $\mathcal{F}^{\lambda}$ consists of the term $\mathcal{F}^{\lambda}_{\mathcal{T}}$, defined as
\be \label{f_t}
\mathcal{F}^{\lambda}_{\mathcal{T}}  = \left(-p+L_M\right)h^\lambda_\nu\nabla^\nu \log{f_2}\,,
\ee
and the hyperforce $\mathcal{F}^\lambda_\mathcal{H}$ is given by
\begin{eqnarray}
\mathcal{S}_\mathcal{H} & = &  -\frac{2}{\sqrt{-g}}u^\nu\Big[\nabla_\alpha \nabla_\beta H_\nu{}^{\alpha\beta} 
%\nonumber \\
%& + & 
\frac{1}{f_2}\lp\nabla_\alpha\nabla_\beta f_2\rp H_\nu{}^{\alpha\beta} + \frac{1}{f_2}f_{2,(\alpha}\nabla_{\beta)}H_\nu{}^{\alpha\beta}\Big]\,.
\end{eqnarray}
It is interesting that the extra force (\ref{f_t}) vanishes identically for a perfect fluid if we adopt the Lagrangian prescription $L_M=p$.

\section{Cosmological applications}\label{secIII}
%\subsection{Cosmological applications}

In the framework of standard Friedman-Robertson-Walker (FRW) geometry, we have $Q=6H^{2}$.
%\begin{equation}  \label{21}
%Q=6H^{2},
%\end{equation}
%
Introducing the effective energy density $\rho _{\rm eff}$ and effective pressure $p_{\rm eff}$ of
the cosmological fluid, defined as
\begin{equation}
\rho_{\rm eff}=-\frac{f_{2}}{2F}\left( \rho -\frac{f_{1}}{2f_{2}}\right) ,
\qquad
p_{\rm eff}=\frac{2\dot{F}}{F}H-\frac{f_{2}}{2F}\left( \rho +2p+\frac{f_{1}}{2f_{2}}\right)\,,
\end{equation}
%\begin{equation}
%p_{\rm eff}=\frac{2\dot{F}}{F}H-\frac{f_{2}}{2F}\left( \rho +2p+\frac{f_{1}}%{2f_{2}}\right) ,
%\end{equation}%
we can write the gravitational field equations in a form similar to the
Friedmann equations of GR as
\begin{eqnarray}
3H^{2}=\rho_{\rm eff},  \qquad
2\dot{H}+3H^{2}=-p_{\rm eff}.  \label{f7}
\end{eqnarray}

To describe cosmological evolution, and the possible transition to
an accelerated phase, we also introduce the parameter $w$ of the dark energy
equation of state, defined as
\begin{equation}
w=\frac{p_{\rm eff}}{\rho_{\rm eff}}=\frac{-4\dot{F}H+f_{2}\left( \rho +2p+\frac{f_{1}%
}{2f_{2}}\right) }{f_{2}\left( \rho -\frac{f_{1}}{2f_{2}}\right) }.
\label{w}
\end{equation}

The deceleration parameter can be written as
\begin{equation}
q=\frac{1}{2}\left( 1+3w\right) = 2 + \frac{3\lp 4\dot{F}H - f_1- 2f_2 p \rp}{f_1-2f_2\rho}\,.
\end{equation}

Thus, as a first step in this direction we have obtained the generalized Friedmann equations describing the cosmological evolution in flat FRW type geometry. The coupling between matter and the $Q$ field introduces two types of corrections. The first is the presence of a term of the form $f_2/2F$ multiplying the components of the energy-momentum tensor (energy density and pressure) in both Friedmann equations. Secondly, an additive term of the form $f_1/4F$ also appears in the generalized Friedmann equations. The basic equations describing the cosmological dynamics can then be reformulated in terms of an effective energy density and pressure, which both depend on the standard components of the energy-momentum tensor, and on the functions $f_i(Q)$, $i=1,2$, and on $F$.
In the vacuum case $\rho =p=0$, the deceleration parameter takes the form $q=-1+12\dot{F}H/f_1$, showing that, depending on the mathematical forms of the coupling functions, a large number of cosmological evolutionary scenarios can be obtained. Generally, we have shown explicitly that for late times, the Universe attains an exponentially accelerating de Sitter phase\cite{Harko:2018gxr}.

\section{Conclusions}\label{secIV}
%\subsection{Summary and Conclusion}

In this work, we have explored an extension of the symmetric teleparallel gravity, by considering a new class of theories where the nonmetricity $Q$ is coupled nonminimally to the matter Lagrangian, in the framework of the metric-affine formalism. As in the standard curvature-matter couplings, this nonminimal $Q$-matter coupling entails the nonconservation of the energy-momentum tensor, and consequently the appearance of an extra force. Thus, in summary, we have established the theoretical consistency and motivations on these extensions of $f(Q)$ family of theories. Furthermore, we considered cosmological applications, in which the presented approach provides gravitational alternatives to dark energy. As future avenues of research, one should aim in characterizing the phenomenology predicted by these theories with a nonmetricity-matter coupling, in order to find constraints arising from observations.

\section*{Acknowledgments}
FSNL acknowledges funding by the Funda\c{c}\~ao para a Ci\^encia e a Tecnologia (FCT, Portugal) through an Investigador FCT Research contract No.~IF/00859/2012. GJO is funded by the Ramon y Cajal contract RYC-2013-13019 (Spain). DRG is funded by the FCT postdoctoral fellowship No.~SFRH/BPD/102958/2014. FSNL and DRG also acknowledge funding from the research grants UID/FIS/04434/2013, No.~PEst-OE/FIS/UI2751/2014 and No. PTDC/FIS-OUT/29048/2017. This work is supported by the Spanish projects FIS2014-57387-C3-1-P,  FIS2017-84440-C2-1-P (AEI/FEDER, EU), the project H2020-MSCA-RISE-2017 Grant FunFiCO-777740, the project SEJI/2017/042 (Generalitat Valenciana), the Consolider Program CPANPHY-1205388, and the Severo Ochoa grant SEV-2014-0398 (Spain).

\end{document}